\documentclass[smallcondensed,natbib]{svjour3} 
\smartqed  
\usepackage{graphicx}
\usepackage{mathptmx}      
\usepackage{latexsym}
\begin{document}

\title{An Overview of the 13:8 Mean Motion Resonance between Venus and Earth}


\subtitle{}


\author{
        {\'A}. Bazs{\'o} \and V. Eybl \and R. Dvorak \and E. Pilat-Lohinger \and Ch. Lhotka
}


\institute{
           Institute of Astronomy, University of Vienna,
           T{\"u}rkenschanzstr. 17, A-1180 Wien, Austria\\
           \email{bazso@astro.univie.ac.at}
}

\date{Received: date / Accepted: date}

\maketitle

\begin{abstract}
It is known since the seminal study of \citet{Las1989} that the inner planetary system
is chaotic with respect to its orbits and even escapes are not impossible,
although in time scales of billions of years. The aim of this investigation
is to locate the orbits of Venus and Earth in phase space, respectively to see
how close their orbits are to chaotic motion which would lead to unstable orbits
for the inner planets on much shorter time scales.
Therefore we did numerical experiments in different dynamical models with
different initial conditions -- on one hand the couple Venus-Earth was set
close to different mean motion resonances (MMR),
and on the other hand Venus' orbital eccentricity (or inclination)
was set to values as large as $e=0.36$ ($i=40^{\circ}$).
The couple Venus-Earth is almost exactly in the 13:8 mean motion resonance.
The stronger acting 8:5 MMR inside, and the 5:3 MMR outside the 13:8 resonance
are within a small shift in the Earth's semimajor axis
(only 1.5 percent). Especially Mercury is strongly affected by relatively small
changes in eccentricity and/or inclination of Venus in these resonances. Even
escapes for the innermost planet are possible which may happen quite rapidly.

\end{abstract}

\section{Introduction}

Mean Motion Resonances are essential in Solar System Dynamics not only for the
planets but also for the motion of the asteroids. The appearance of chaotic
motion in the asteroid belt detected by \citet{Wis1981} -- the 3:1 MMR of an
asteroid with Jupiter --
was the first one discovered after the seminal discovery by \citet{Hen1964}
for galactic dynamics. But the structure of the asteroid belt is
also created by the secular resonances (SR), where the motion of the perihelia and
the nodes can be in resonance for different planets in combination with the
asteroid's perihelion respectively node. For the planets it was found by
Laskar in different papers (\citet{Las1988,Las1990,Las1996}) that especially the
Inner Solar System (ISS) is in a chaotic state. The appearance of SR may even shift
the orbit of Mars out from a stable orbit into an unstable one with high
probability when one does not take into account relativity.
The large planets seem to be in quite a 'safe' region, although many MMR are
acting: e.g. the 5:2 MMR between Jupiter and Saturn. Recent
investigations by \citet{Mic2001,Gal2006} show how close their motion is to a chaotic state with
severe implications for the other planets.

\begin{figure}
\begin{center}
\includegraphics[width=3.0in,angle=270]{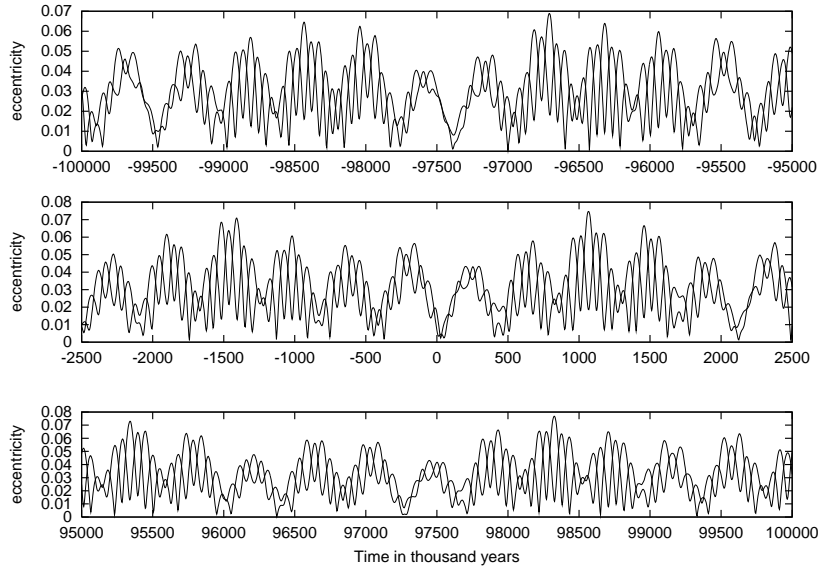}
\caption{The coupling in the eccentricities of Venus and Earth; from -100 to 95
  Myrs (upper graph), from -2.5 to 2.5 Myrs (middle graph) and from 95 to
  100 Myrs (lower graph), after \citet{Dvo2003}.}
\label{Fig1}
\end{center}
\end{figure}

\begin{figure}
\begin{center}
\includegraphics[width=4.5in,angle=0]{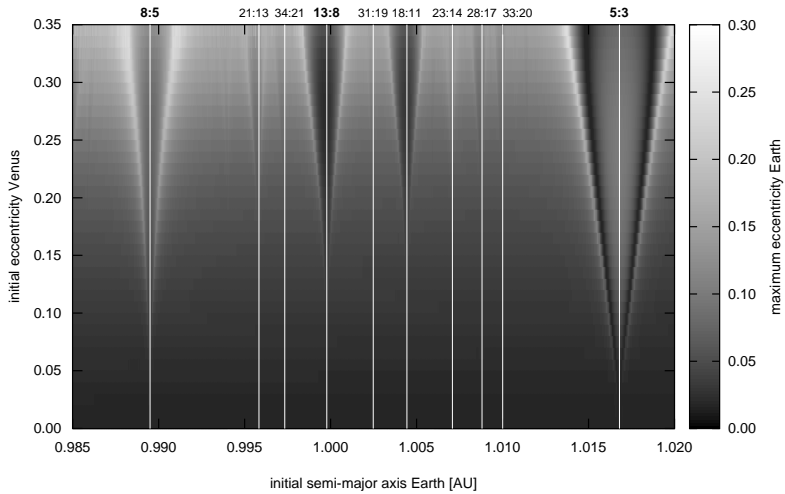}
\caption{The MMRs between Venus and the Earth; along the x-axis the initial semimajor
  axis of the Earth is changed, along the y-axis the initial values for
  the eccentricity of Venus are varied. The greytone indicates the maximum
  eccentricity, the vertical white lines show the position of the MMR up to
  high orders.}
\label{Fig2}
\end{center}
\end{figure}

Because the orbits of Venus and Earth are quite close to the 13:8
MMR\footnote{the periods of Venus and Earth are almost in the ratio 13:8}
and they are strongly coupled with respect to the inclinations and
eccentricities (Fig. \ref{Fig1}) -- comparable to
Jupiter and Saturn -- we investigated the phase space close to this high order
resonance. The order of a resonance is defined as the difference $q$
when we characterize a MMR by $p:(p+q)$. The value $q$ is connected to the
exponent in front of the fourier term in the expansion of the disturbing function; the higher
the exponent in a small quantity in front of the Fourier term is, the less strong
is the influence on the dynamics, unless there is a resonance acting
(e.g. Murray and Dermott). Thus for the 13:8 MMR we have to deal with the
order 5. It is interesting to note that a relatively small shift
of the Earth to a smaller
semimajor axis would bring the couple Venus--Earth into the 8:5 MMR (order 3); a shift
to a larger semimajor axis would bring both planets into the 5:3 MMR (order 2) (Fig. \ref{Fig2}). In this plot the results of the computations in the simple three body problem
Sun-Venus-Earth are presented to show the location of the resonances and also
their shape.

Besides extensive numerical integration of the motions of the planets in different dynamical
models, we also used chaos indicators and analysed the frequencies involved. Furthermore simplified
mapping models were constructed to understand the structure of phase space where Venus and
Earth are embedded.

The content of the paper is follows: after a careful test of the
dynamical model which we used (section 2) we roughly describe
the three MMR mentioned above and show the structure determined
with the aid of a chaos indicator (section 3). We then report on the results of
different numerical experiments where we show the dependency of the stability
of the inner planetary system on the eccentricities and the inclinations of
Venus (section 4). In the conclusions we discuss the possible consequences
for the stability of the inner planetary system. In an appendix first results of an
applied mapping for motions in MMRs are shown.

\section{The dynamical models and the numerical setup}

To choose a quite realistic model we first did integrations with the actual
initial conditions of Venus and Earth in different dynamical models, using
a numerical integration scheme based on Lie-series (\citet{Han1984, Del1984, Lic1984}):

\begin{description}
  \item[\bf A] the Inner Solar System (ISS, 5-bodies)
  \item[\bf B] the ISS + Jupiter (6-bodies)
  \item[\bf C] the ISS + Jupiter + Saturn (7-bodies)
  \item[\bf D] the ISS + Jupiter + Saturn + Uranus (8-bodies)
  \item[\bf E] the ISS + outer Solar system (9-bodies)
\end{description}

\begin{figure}
\begin{center}
\includegraphics[width=3.0in,angle=270]{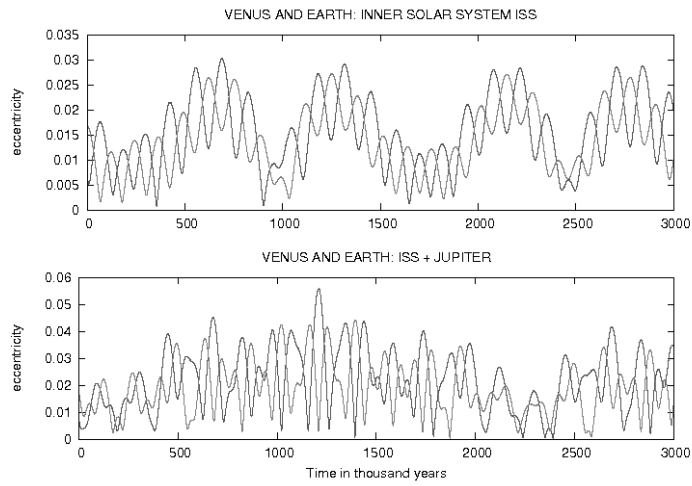}
\caption{Coupling in the eccentricities of Venus and Earth for the model
{\bf A} (upper graph) and in the model {\bf B} for 3 million years.}
\label{Fig3}
\end{center}
\end{figure}

\begin{figure}
\begin{center}
\includegraphics[width=3.0in,angle=270]{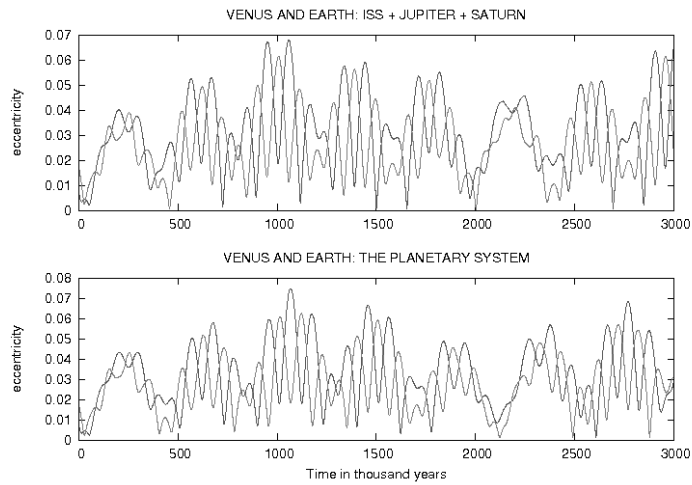}
\caption{Coupling in the eccentricities of Venus and Earth for the model
{\bf C} (upper graph) and in the model {\bf E} for 3 million years.}
\label{Fig4}
\end{center}
\end{figure}

The different behaviour between the models consisting only of the inner planets
({\bf A}) and the one with  Jupiter ({\bf B}) is visible in Fig. \ref{Fig3}: the very
regular variation limited by a maximum value of $e=0.03$ (upper graph) -- is completely destroyed by the
perturbation of Jupiter (Fig. \ref{Fig3}: lower graph). One can see three
distinct sections:
(i) sometimes the eccentricities almost do not change
for up to $10^5$ years (e.g. between $2.25$ and $2.35 \times 10^6$ years for Venus and
$2.55$ and $2.65 \times 10^6$ years for the Earth); (ii) sometimes the changes are in
antiphase (like in the upper graph of Fig. \ref{Fig3}); and (iii) sometimes they are in phase.

Including now Saturn as perturber in model {\bf C} changes the picture a lot: the development
of the eccentricities is always in antiphase, reaches even larger maximum
values, but looks 'smoother' for both planets compared to model {\bf B} (Fig. \ref{Fig4}
upper graph). Now by comparing the dynamical evolution of Venus and Earth
including all planets of the Solar System (model {\bf E}, Fig. \ref{Fig4} bottom)
with {\bf C} it is interesting to note that for the first $10^6$ years even quantitatively the
behaviour is very similar. Later (up to $3 \times 10^6$ years) at least qualitatively both
models show an analogous behaviour\footnote{Model {\bf D}, not shown here, is
not significantly better than {\bf C}.}.

\begin{figure}
\begin{center}
\includegraphics[width=3.0in,angle=270]{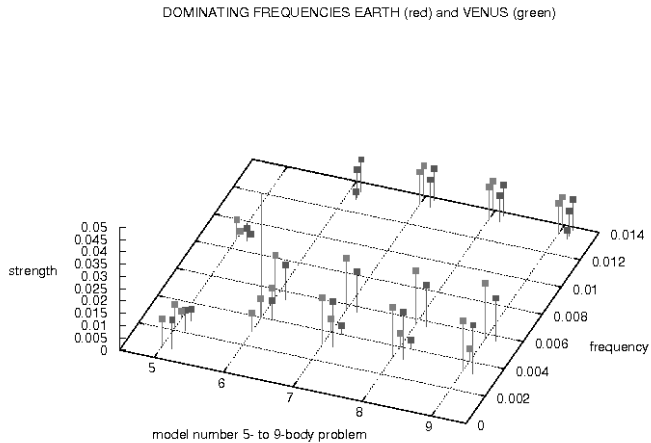}
\caption{Comparison of the 5 dominating frequencies in the dynamics of Venus
  and Earth for the five different models {\bf A} to {\bf E}. On the x-axis the
  number of bodies is given (5 to 9), on the y-axis the frequencies (in arcsec per year) and in
  z-direction the amplitude is given. Always on the left is the frequency of
  Earth, on the right the frequency of Venus.The numbers on the x-axis show
the number of bodies in the respective models; e.g. '5' means that the 5-body problem
was used for the numerical integration.}
\label{Fig5}
\end{center}
\end{figure}

In addition to the former computations we compare for the five
different models the dominating frequencies,
which were derived by the program {\sc SigSpec} (\citet{Ree2007}\footnote{This code is a
new method in time series analysis
and it is the first technique to rely on an analytic solution for the
probability distribution produced by white noise in an amplitude
spectrum (Discrete Fourier Transform). Returning the probability that
an amplitude level is due to a random process, it provides an
objective and unbiased estimator for the significance of a detected
signal.}). {\sc SigSpec} incorporates the frequency and phase of the
signal appropriately and takes into account the properties of the
time-domain sampling which provides an unprecedentedly accurate peak
detection (\citet{Kal2008}). Although the present
investigations deal with equidistantly sampled time series, the
benefit of frequency- and phase-resolved statistics and the capability
of full automation are indispensible also in this application.

The data used were the orbital element $k=e \cos \varpi$ of the Earth and
of Venus in
all mentioned dynamical models. It is evident that the
model {\bf A} gives quite a wrong picture of the
involved resonances.
To include only Jupiter is already a much better model and taking into account
the great inequality between Jupiter and Saturn {\bf C} (x-axes '7') is a very good
description of the motions of Earth and Venus. The differences between the
models {\bf C} and {\bf E} are not significant as we also noticed from the direct inspection
of the eccentricity plots. Consequently as compromise between long CPU time and
precision of the model it is clear that {\bf C} is a good choice.

\begin{figure}
\begin{center}
\includegraphics[width=3.0in,angle=270]{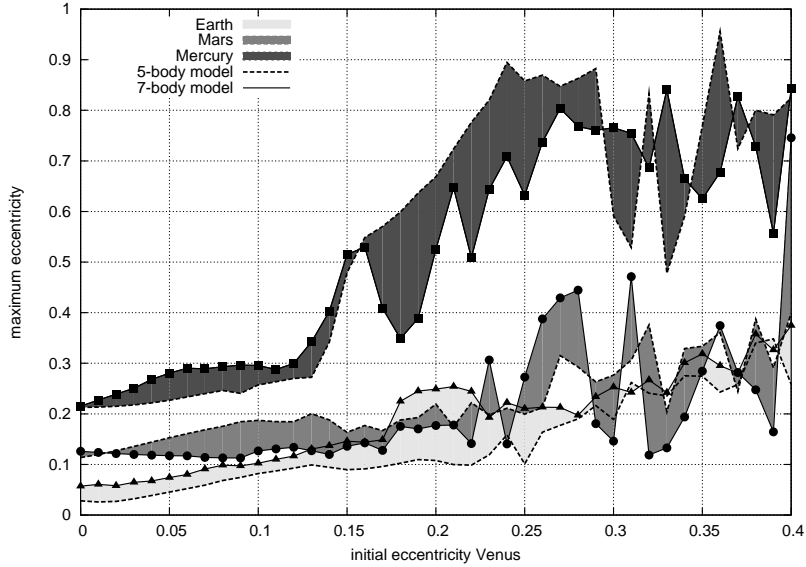}
\caption{Results of integrations for $10^{6}$ years with fixed orbital
  elements in the 13:8 MMR: the differences in maximum eccentricity of
  each individual planet in 2 models ({\bf A}, dashed lines, {\bf C} full lines)
  are plotted (y-axis) for initially different inclinations of Venus (x-axis).
  Light grey: Earth, grey: Mars, dark grey: Mercury.}
\label{Fig6}
\end{center}
\end{figure}

One additional test was made by checking the maximum eccentricities of the
orbits ($e_{max}$) of the inner planets when changing the initial eccentricity of Venus.
The respective integrations for $10^{6}$ years shown in Fig. \ref{Fig6} point out
the differences in $e_{max}$ of each individual planet in the 5 and 7-body models ({\bf A}
and {\bf C}) for the 13:8 MMR.
In contrast to the other planets one can see that the Earth (light grey)
has almost always a higher maximum eccentricity due to the
influence of Jupiter and Saturn than without these two
planets. For Mercury and Mars no clear dependence is observable, the maximum
values of their eccentricities change with the initial inclination of Venus.
It is remarkable that Mercury achieves very high
eccentricities, and eventually escapes for longer integrations (see next chapter).

\section{Description and analysis of the 8:5, 13:8 and 5:3 MMR}

In Fig. \ref{Fig2} we show a detailed graph of the neighborhood of the 13:8 MMR in the
planar 3-body problem Sun-Venus-Earth: in addition to the 13:8, the 8:5 and the
5:3 MMR many high order MMRs are visible. To derive a simple picture of the
phase space structure close to the 13:8 MMR only two initial parameters
were changed: on the x-axis the semimajor axis of the Earth was varied and on the y-axis
the initial eccentricity of the perturbing planet Venus is plotted. In this
graph, where the maximum eccentricity of an Earth-orbit during the integration
time is shown on a grey scale, one already can see the inequality of the perturbations
acting in the resonances.

\begin{itemize}

\item The 8:5 MMR (order 3, $a=0.989501$ AU)\footnote{with $a_{Venus}= 0.723330$
AU} has a relatively narrow triangular structure with
highly perturbed wings down to $e_{Venus}=0.12$ on both sides, whereas in the middle of the resonance the
orbits are very regular with small eccentricities.

\item The 13:8 MMR (order 5, at $a=0.999782$ AU) is the weakest of the three MMR we are studying.
The wings on both sides of the resonance go down to $e_{Venus}=0.15$. Again in
the middle of the resonance -- here broader than in the 8:5 resonance -- the
orbits are only marginally perturbed; the eccentricities stay well below
0.075\footnote{The Earth with a semimajor axis $a=1$ is just a little outside of this MMR.}.

\item The 5:3 MMR (order 2, $a=1.016799$ AU) is by far the strongest one: it is quite broad from
  large eccentricities of Venus on down to small eccentricities like $e=0.05$.
It shows a big 'quiet' central region and in the wings the Earth's orbit
suffers from relatively large eccentricities $e_{Earth} \sim 0.3$
\end{itemize}

Another high order MMR, the 18:11 resonance, is visible close to the 13:8
resonance, which we will not discuss here because its action is evidently
weaker (order 7).

To use the results of direct numerical
integrations one has to carry out orbital computations over a very long time. In order to save
computation time it is advisable to use a so-called chaos indicator, that
shows the state of motion quite fast and allows us to reduce the integration
time significantly. Therefore, we used the {\sc Fast Lyapunov indicator (FLI)}
(see \citet{Fro1997}), which is a quite fast tool to distinguish between regular
and chaotic motion. According to the definition --
where $\Psi$ is the length of the largest tangent vector:

\begin{eqnarray*}
\psi (t) = \sup_i \|v_i(t)\| \qquad i=1,\dots n
\end{eqnarray*}

\begin{figure}
\begin{center}
\includegraphics[width=3.5in,angle=0]{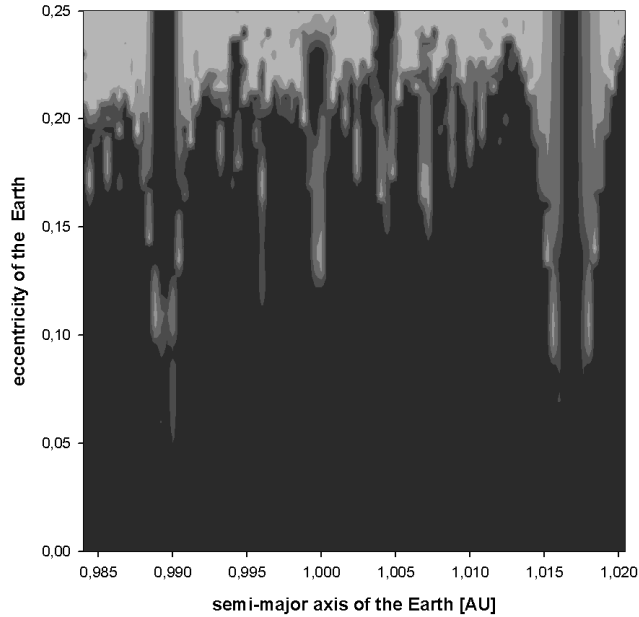}
\caption{FLI map: semimajor axes of the Earth versus different eccentricities of
  Venus. Visible from left to right the 8:5, the 13:8 and the 5:3 MMR. Black
  regions indicate stable orbits.}
\label{Fig7}
\end{center}
\end{figure}

\begin{figure}
\begin{center}
\includegraphics[width=3.5in,angle=0]{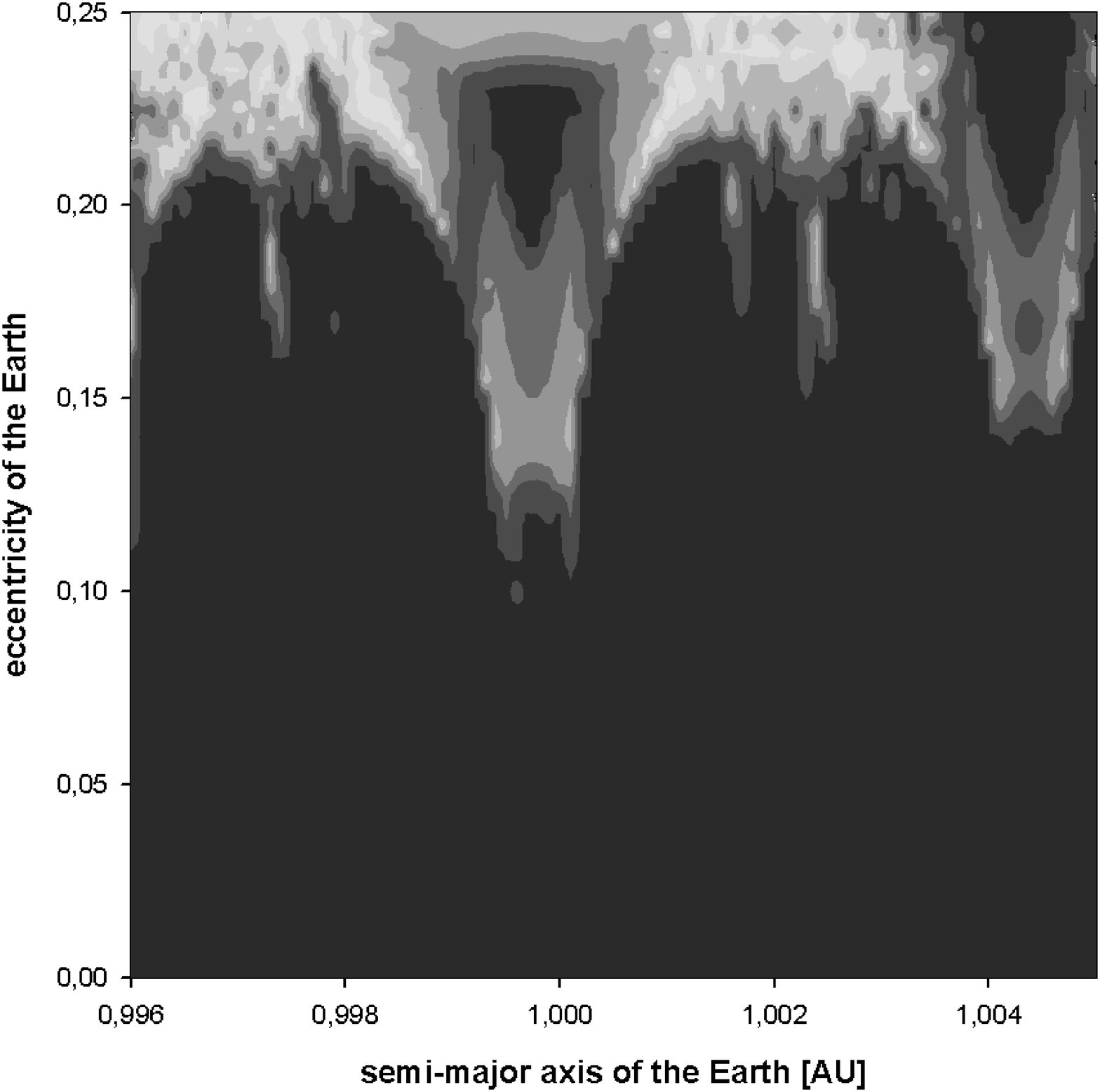}
\caption{Zoom of the 13:8 MMR, captions like in Fig. \ref{Fig7}.}
\label{Fig8}
\end{center}
\end{figure}

($n$ denotes the dimension of the phase space) -- it is obvious that chaotic
orbits can be found very quickly because of the exponential growth of this
vector in the chaotic region. This method has often been applied to
studies of  Extra-solar Planetary Systems (\citet{Loh2008, Schw2009}).
In this study the FLIs were computed for about 50000 yrs.
The resulting stability maps are shown in Figs. \ref{Fig7} and \ref{Fig8}, where the black
region marks the stable motion and grey areas show chaotic regions.
Fig. \ref{Fig7} shows the state of motion from 0.98 AU to 1.020 AU for different
eccentricities of the Earth. One can clearly see perturbations in the stable
region even for low eccentric motion of the Earth due to MMR with Venus. The
three MMRs -- 8:5, 13:8 and 5:3 -- that
are analyzed in this study are clearly visible in Fig. \ref{Fig7}.
A magnification of the 13:8 MMR is shown in Fig. \ref{Fig8}, which indicates
chaotic
behaviour for the position of the Earth if $e_{Earth} > 0.1$. The interesting
fine structures inside the MMR (symmetric with respect to the central line of
the resonance) unveil a central region for $e=0.2$ which still shows regular
motions of the Earth, while close to it chaotic motion is
present (white in Fig. \ref{Fig7}).

\section{Dependence on the initial conditions of Venus}

To check the sensitivity with respect
to different initial conditions of the 'planet twins' Venus--Earth concerning
the stability of the inner Solar System we performed a series of
numerical integrations in the model {\bf C} for $10^7$ years where we checked
the three MMR described before. Two cuts through the exact location of the MMR
have been undertaken, one where we changed the eccentricity of Venus and
another one where we changed its inclination. Both results were analysed
through the simple test of the largest eccentricity of the 4 inner planets
during the integration.

\subsection{Dependence of the Earth's orbit on the eccentricity of Venus}

\begin{figure}
\begin{center}
\includegraphics[width=4.0in,angle=0]{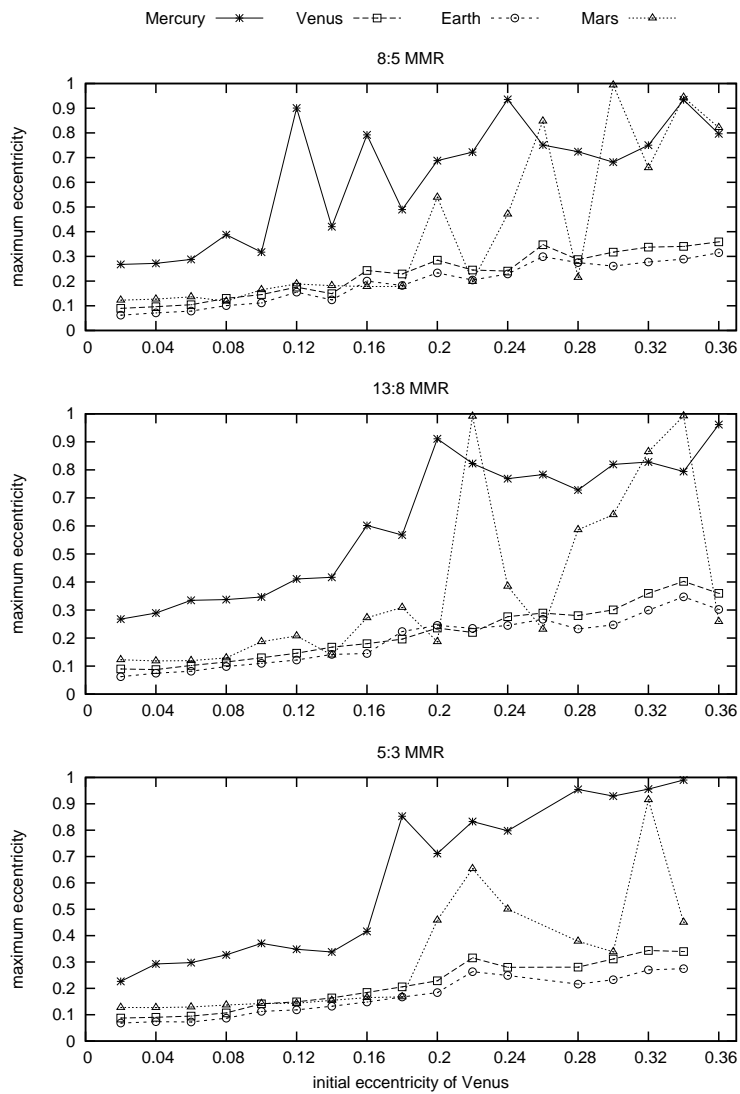}
\caption{$e_{max}$ cuts through the centers of the 8:5, 13:8 and 5:3 MMR (from top to bottom) depending on the eccentricity of the perturbing planet Venus (x-axis).
}
\label{Fig9}
\end{center}
\end{figure}

\begin{itemize}

\item {\bf The 8:5 MMR} (Fig. \ref{Fig9}, upper graph):

The orbits of Venus and the Earth develop qualitatively in the same way, and
no dramatic changes are visible in their orbits. Mercury is the planet which
suffers -- even for relatively small initial eccentricities of Venus ($e_{Venus} \le
0.12$) -- from large eccentricities. For all larger values of
$e_{Venus}$ Mercury is in quite a chaotic orbit. Mars remains longer with
orbital elements which
do not allow close encounters to the Earth, but from $e_{Venus} \ge
0.2$ on its orbit is also destabilized.

\item {\bf 13:8 MMR} (Fig. \ref{Fig9}, middle graph):

Up to $e_{Venus} \leq 0.15$ the inner planetary system seems to be in a
relatively 'quiet' region of phase space; from that on, Mercury and Mars getting
larger and larger eccentricities, the system is destabilized. Venus and Earth
are still developing in quite a similar way without dramatic changes of their orbits.

\item {\bf The 5:3 MMR} (Fig. \ref{Fig9}, lower graph):

The orbit of the Earth suffers more and more from larger and larger
eccentricities depending on the one of Venus very similar to the other two
resonances.
The two planets Mercury and Mars achieve such large eccentricities that encounters with Venus
respectively Earth cause them to leave their orbits. The chaotic behaviour
of the orbits of the inner planetary system is already visible in the FLI
diagram for $e_{Venus} \geq 0.17$ on (compare Fig. \ref{Fig7}).

\end{itemize}

The dynamical behaviour of orbits inside these three resonances is quite
similar; although the 5:3 and 8:3 MMR are of a lower order compared to the 13:8
MMR and one expects stronger perturbations on the orbits of the planets;
in the respective plots this is not visible.

\subsection{Dependence of the Earth's orbit on the inclination of Venus}

\begin{figure}
\begin{center}
\includegraphics[width=4.0in,angle=0]{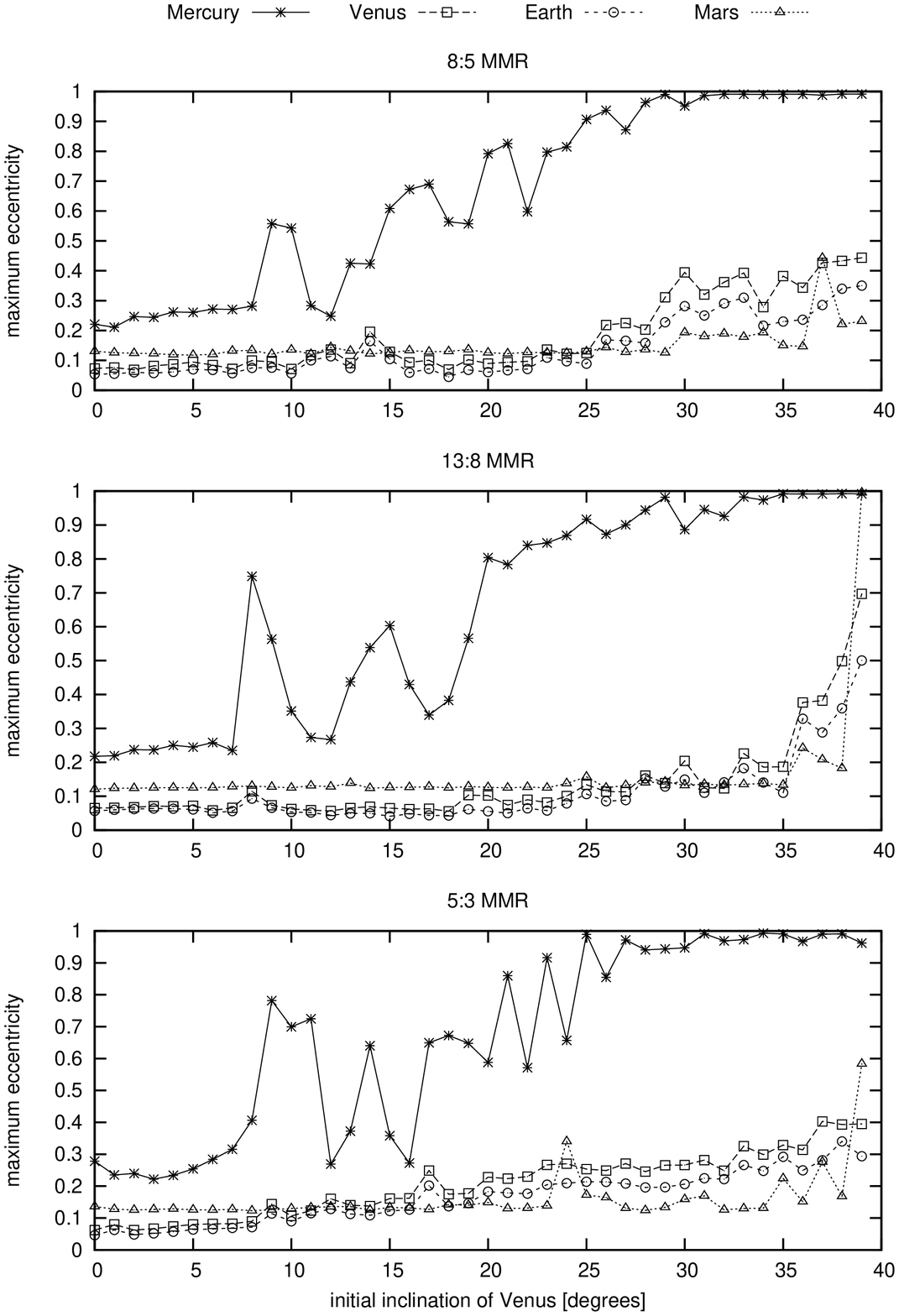}
\caption{$e_{max}$ cuts through the centers of the 8:5, 13:8 and 5:3 MMR (from top to bottom) depending on the inclination of the perturbing planet Venus (x-axis).
}
\label{Fig10}
\end{center}
\end{figure}

\begin{itemize}

\item  {\bf The 8:5 MMR} (Fig. \ref{Fig10}, upper graph):

In this resonance it is evident that Mercury is thrown out from its
orbit already for moderate initial inclinations of Venus, although a small
window of more stable orbits appears for $11^{\circ} \leq i \leq
12^{\circ}$. Earth and Venus
are strongly perturbed: from $i=29^{\circ}$ on eccentricities
up to $e=0.35$ for Venus and $e=0.3$ for the Earth are possible. Mars obviously is
not so much affected by the Earth, only for $i=37^{\circ}$ a maximum value of $e=0.4$ is
reached, which brings him into an Earth-crossing orbit. For longer
integration time again
an escape seems possible. Then, surprisingly, Mars for $i=38^{\circ}$ and $i=39^{\circ}$ (within the
integration time of $10^7$ years!) still stays in a moderatly elliptic orbit.

\item  {\bf The 13:8 MMR} (Fig. \ref{Fig10}, middle graph):

It is quite remarkable that even a small shift of the initial inclination to
$i \sim 8 ^{\circ}$ of
Venus may lead to an escape of Mercury (after $e=0.75$ a subsequent escape
from its orbit because of encounters with Venus is expected). A second 'dangerous'
inclination is visible around $i=15^{\circ}$; then, from $i \ge 20 ^{\circ}$ on, Mercury
would suffer from close approaches to Venus. There seems to be little influence
on the orbits of Venus, Earth and Mars up to $i_{Venus}=35^{\circ}$.

\item  {\bf The 5:3 MMR} (Fig. \ref{Fig10}, lower graph):

The strongest of the three MMR shows in fact a dynamical evolution which is
dominated by the very strong perturbations on Mercury for inclinations of
Venus around $10^{\circ}$ and then, from $15^{\circ}$ on the large eccentricities of
Mercury ($e \geq 0.6$) will always lead to escapes of this planet. Mars is not
affected dramatically in this resonance, Venus and Earth stay within moderate
eccentricities for all initial inclinations of Venus.

\end{itemize}

\section{Conclusions}

In this investigation we explored the environment of the 13:8 MMR, which is very
close to the actual position of Venus and Earth.
After a careful test of a simplified dynamical model of our planetary system
appropriate for our task we discussed the main structures of the three
MMR in the vicinity of the two planets using the results of a study of the
{\sc FLI}.

In another step, using extensive numerical integrations
we changed the ratio of the
semimajor axes Venus--Earth\footnote{Venus' semimajor axis was fixed.}; the
orbit of the Earth was given different
values of $a$ to cover the proximate neighborhood of the MMR where the
twin-planets are in. Then, in two different runs the eccentricity of Venus was
set to eccentricities up to $e = 0.36$ and in the dynamical model, consisting
of the inner Solar System with Jupiter and Saturn, the equations of motion were
integrated for $10^7$ years. The same was done in a second run where we changed
the inclination of Venus up to $i=40^{\circ}$. These two experiments were
undertaken for each of the three MMR 8:5, 13:8 and 5:3 and the results analysed with
respect to the maximum eccentricties of the inner planets. It turned out that
only Mercury is the planet which would suffer from such perturbations, that it
would achieve very large values of its eccentricity and thus would be thrown out from its current
orbit. According to our results Mars is not as strongly affected by a larger
inclination or eccentricity of Venus as Mercury; this can be understood by the presence
of Jupiter and Saturn keeping this planet almost in its original plane of
motion. Venus and also the Earth are only affected for larger initial
inclinations of Venus, but the planets stay more or less in their orbits
although sometimes achieving eccentricities as large as $e=0.3$.

In an attempt to understand the dynamical structure of the resonances
a mapping approach (appendix) was constructed in the simple elliptic restricted three body
problem Sun--Venus and massless Earth for the 3:1 MMR.
The same approach was then used to construct the surface of
section of the 5:3 MMR. Until now we failed to get a similar graph for the
other two resonances which ask for a development of the Hamiltionian up to
order 3 and order 5, but we work on it.

As a final statement we may say that according to our results although the
couple of Venus and Earth is close to a MMR we do not expect big changes of
the orbits of the inner planets unless Venus will come into an orbit of an
inclination of about 7 degrees or will achieve an eccentricity of 0.2. Then, in
fact, Mercury would be on an unstable orbit which could finally throw it either
into the Sun or far out into the main belt of asteroids!

\section{Appendix: A Mapping Model}

\begin{figure}
\begin{center}
\includegraphics[width=2.3in,angle=0]{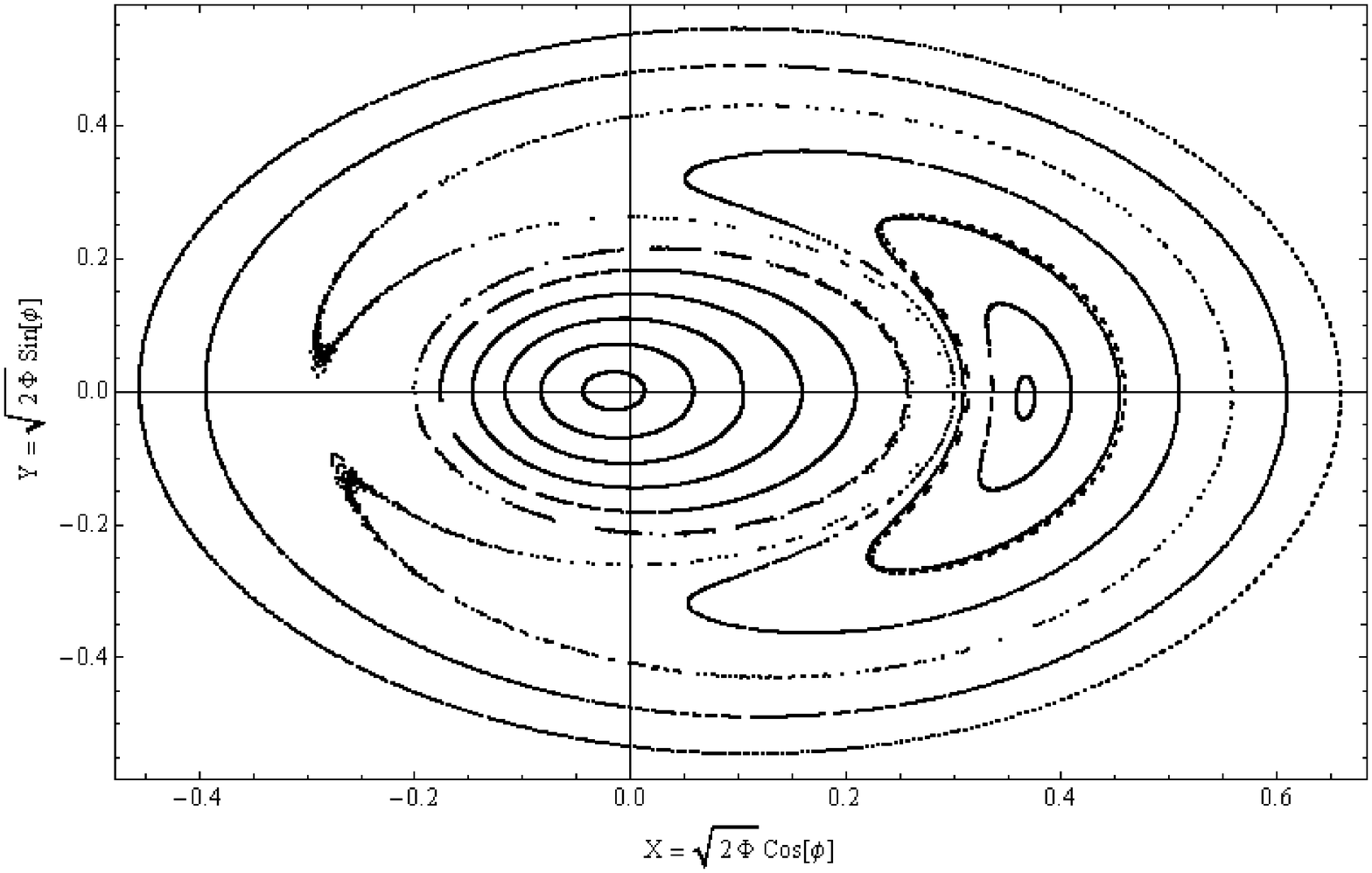}
\includegraphics[width=2.3in,angle=0]{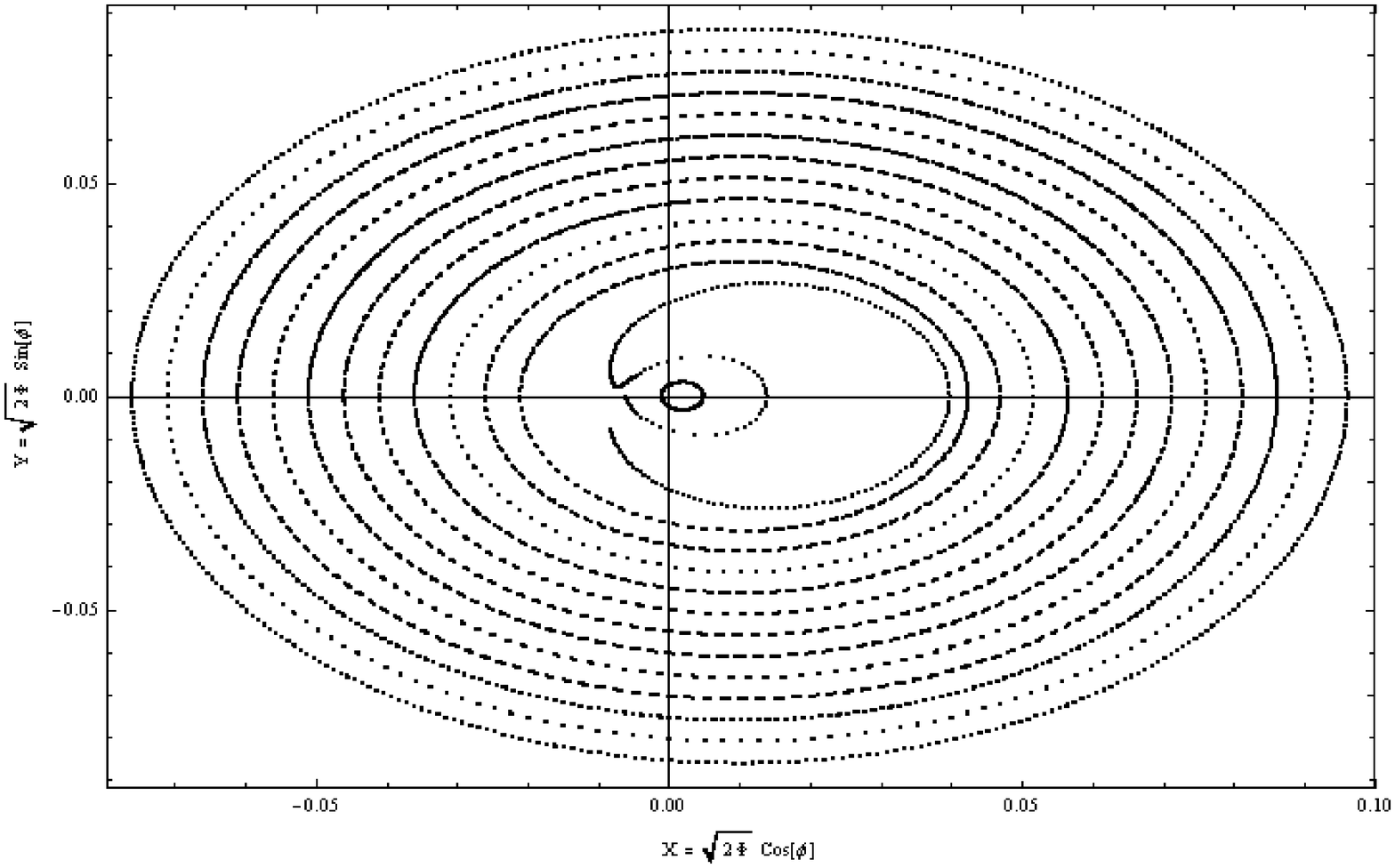}
\caption{ Phase space portrait of the 3:1 resonance
(left) and 5:3 resonance (right) of a massless body with Earth.}
\label{Fig11}
\end{center}
\end{figure}

Mapping methods are very efficient analytical tools for studying the dynamics
of mean motion resonances.  Their main advantage is the fact, that a mapping
is much faster in regard to computing time,
while also being more exact than numerical integrations in the same model. In this work, we
followed the approach used in the works of \citet{Had1993, Tsi2007, Lho2008},
which has been found very successful already in the study of the 1:1
resonance. The mapping allows us to investigate the phase space structures of $p:(p+q)$ MMR
in a semi-analytical way. The Hamiltonian of the three body problem in the
exact location of the chosen resonance is used to derive equations, which
provide a set of rules for the propagation of a trajectory in phase space.
The Hamiltonian in Keplerian elements $(a, e, \omega, \Omega)$
describing the motion of the perturbed body is averaged over the orbital period of the
perturbing body and then transformed to action and angle variables
 $(X, Y, \psi, J)$ in phase space by canonical transformations. The mapping equations then
provide iteration rules for each of these variables with a time step equaling the
orbital period of the perturber. Being based on the exact equations of motion,
the results obtained from the mapping equations provide the most accurate
insight in the dynamics of the chosen system.

To test the applicability
of the method, the 3:1 resonance of a massless body with Jupiter in the planar
elliptic restricted three body problem has been analyzed, which yielded very
good results. The same method was used for the 5:3 resonance of a massless particle with Earth,
again in the planar ER3BP. These two resonances are of order two
and therefore more easily computable than the 8:5 and the 13:8 resonance,
In Fig. \ref{Fig11} (left graph) the phase space portrait for the 3:1 mean motion resonance of a
test body with Jupiter is shown. The same method was applied to the 5:3 MMR
with Earth.
Up to now, we have been studying only low-order resonances in the planar elliptic
restricted three body problem using the described method with good
results. In a next step we will expand this work to the investigation
of the 8:5 MMR and especially the 13:8 MMR, which was the task of this study.

\section{Acknowledgements}

A. Bazso, V. Eybl and Ch. Lhotka appreciate the financial help of the FWF
project P 18930-N16; E. Pilat-Lohinger needs to thank the FWF project P-19569-N16.

%
%
%


\begin{thebibliography}{}


\bibitem[Delva (1984)]{Del1984}
{Delva}, M., "{Integration of the elliptic restricted three-body problem with Lie series}", {Celest. Mech.}, 34, {145-154}, (1984)




\bibitem[Dvorak and S{\"u}li (2002)]{Dvo2002}
{Dvorak}, R. and {S{\"u}li}, {\'A}., "{On the Stability of the Terrestrial
Planets as Models for Exosolar Planetary Systems}", {Celest. Mech. Dyn.
Astron.}, 83, {77-95}, (2002)


\bibitem[Dvorak and Gamsj{\"a}ger (2003)]{Dvo2003}
{Dvorak}, R. and {Gamsj{\"a}ger}, C., "{A New Determination of the Basic Frequencies in Planetary Motion}", {Proceedings of the 3rd Austrian-Hungarian Workshop on Trojans and related Topics}, {49-58}. {F. Freistetter, R. Dvorak and B. {\'E}rdi) (Eds.)}, (2003)


\bibitem[Froeschl{\'e} et al (1997)]{Fro1997}
{Froeschl{\'e}}, C. and {Lega}, E. and {Gonczi}, R., "{Fast Lyapunov
Indicators. Application to Asteroidal Motion}", {Celest. Mech. Dyn.
Astron.}, 67, {41-62}, (1997)


\bibitem[Gallardo (2006)]{Gal2006}
{Gallardo}, T., "{Atlas of the mean motion resonances in the Solar
System}", {Icarus}, 184, {29-38}, (2006)


\bibitem[Hadjidemetriou (1993)]{Had1993}
{Hadjidemetriou}, J.~D., "{Asteroid motion near the 3:1 resonance}",
{Celest. Mech. Dyn. Astron.}, 56, {563-599}, (1993)


\bibitem[Hanslmeier and Dvorak (1984)]{Han1984}
{Hanslmeier}, A. and {Dvorak}, R., "{Numerical Integration with Lie
Series}", {A\&A}, 132, {203-+}, (1984)


\bibitem[Henon and Heiles (1964)]{Hen1964}
{Henon}, M. and {Heiles}, C., "{The applicability of the third integral of
motion: Some numerical experiments}", {AJ}, 69, {73-+}, (1964)


\bibitem[Kallinger et al (2008)]{Kal2008}
{Kallinger}, T. and {Reegen}, P. and {Weiss}, W.~W., "{A heuristic
derivation of the uncertainty for frequency determination in time series
data}", {A\&A}, 481, {571-574}, (2008)


\bibitem[Laskar (1988)]{Las1988}
{Laskar}, J., "{Secular evolution of the solar system over 10 million
years}", {A\&A}, 198, {341-362}, (1988)


\bibitem[Laskar (1989)]{Las1989}
{Laskar}, J., "{A numerical experiment on the chaotic behaviour of the solar system}", {Nature}, 338, {237-+}, (1989)


\bibitem[Laskar (1990)]{Las1990}
{Laskar}, J., "{The chaotic motion of the solar system - A numerical estimate of the size of the chaotic zones}", {Icarus}, 88, {266-291}, (1990)


\bibitem[Laskar (1994)]{Las1994}
{Laskar}, J., "{Large-scale chaos in the solar system}", {A\&A}, 287, {L9-L12}, (1994)


\bibitem[Laskar (1996)]{Las1996}
{Laskar}, J., "{Large Scale Chaos and Marginal Stability in the Solar System}", {Celest. Mech. Dynam. Astron.}, 64, {115-162}, (1996)


\bibitem[Laskar (1997)]{Las1997}
{Laskar}, J., "{Large scale chaos and the spacing of the inner planets.}",
{A\&A}, 317, {L75-L78}, (1997)


\bibitem[Lhotka et al (2008)]{Lho2008}
{Lhotka}, C. and {Efthymiopoulos}, C. and {Dvorak}, R., "{Nekhoroshev
stability at $L_{4}$ or $L_{5}$ in the elliptic-restricted three-body
problem - application to Trojan asteroids}", {MNRAS}, 384, {1165-1177},
(2008)


\bibitem[Lichtenegger (1984)]{Lic1984}
{Lichtenegger}, H., "{The dynamics of bodies with variable masses}", {Celest. Mech.}, 34, {357-368}, (1984)


\bibitem[Lohinger et al (2008)]{Loh2008}
{Pilat-Lohinger}, E. and {Robutel}, P. and {S{\"u}li}, {\'A}. and {Freistetter}, F., "{On the stability of Earth-like planets in multi-planet systems}", {Celest. Mech. Dyn. Astron.}, 102, {83-95}, (2008)


\bibitem[Michtchenko and Ferraz-Mello (2001)]{Mic2001}
{Michtchenko}, T.~A. and {Ferraz-Mello}, S., "{Modeling the 5 : 2
Mean-Motion Resonance in the Jupiter-Saturn Planetary System}", {Icarus},
149, {357-374}, (2001)


\bibitem[Reegen (2007)]{Ree2007}
{Reegen}, P., "{SigSpec. I. Frequency- and phase-resolved significance in
Fourier space}", {A\&A}, 467, {1353-1371}, (2007)


\bibitem[Schwarz et al (2009)]{Schw2009}
{Schwarz}, R. and {S{\"u}li}, {\'A}. and {Dvorak}, R. and {Pilat-Lohinger}, E., "{Stability of Trojan planets in multi-planetary systems. Stability of Trojan planets in different dynamical systems}", 104, {69-84}, (2009)


\bibitem[Tsiganis (2007)]{Tsi2007}
{Tsiganis}, K., "Chaotic Diffusion of Asteroids", {Lect. Notes Phys.},
Springer (Berlin Heidelberg), 729, {111-150}. D. {Benest}, C.
{Froeschl{\'e}}, and E. {Lega} (Eds.), (2007)


\bibitem[Wisdom (1981)]{Wis1981}
{Wisdom}, J.~L., "{The origin of the Kirkwood gaps: A mapping for
asteroidal motion near the 3/1 commensurability.}". {California Inst.~of
Tech.}, Pasadena, (1981)


\end{thebibliography}
\end{document}